\documentclass[prx,floatfix,twocolumn,superscriptaddress]{revtex4-1}

\usepackage[english]{babel}
\usepackage{amsmath}
\usepackage{graphicx}
\usepackage{float}
\usepackage{bm}
\usepackage{multirow}
\usepackage{dcolumn}
\usepackage{hyperref}
\usepackage{color}

\newcommand{\icm}{\ensuremath{~\textrm{cm}^{-1}}}

\begin{document}

\title{\textbf{Unraveling the origin of Kondo-like behavior in the 3$d$-electron heavy-fermion compound YFe$_{2}$Ge$_{2}$}}

\author{Bing Xu}
\email[]{bingxu@iphy.ac.cn}
\thanks{These authors contributed equally to this work.}
\affiliation{Beijing National Laboratory for Condensed Matter Physics, Institute of Physics, Chinese Academy of Sciences, P.O. Box 603, Beijing 100190, China}
\affiliation{School of Physical Sciences, University of Chinese Academy of Sciences, Beijing 100049, China}

\author{Rui Liu}
\thanks{These authors contributed equally to this work.}
\affiliation{Center for Advanced Quantum Studies and Department of Physics, Beijing Normal University, Beijing 100875, China}

\author{Hongliang Wo}
\affiliation{State Key Laboratory of Surface Physics and Department of Physics, Fudan University, Shanghai 200433, China}

\author{Zhiyu Liao}
\author{Shaohui Yi}
\affiliation{Beijing National Laboratory for Condensed Matter Physics, Institute of Physics, Chinese Academy of Sciences, P.O. Box 603, Beijing 100190, China}
\affiliation{School of Physical Sciences, University of Chinese Academy of Sciences, Beijing 100049, China}

\author{Chunhong Li}
\affiliation{Beijing National Laboratory for Condensed Matter Physics, Institute of Physics, Chinese Academy of Sciences, P.O. Box 603, Beijing 100190, China}

\author{Jun Zhao}
\affiliation{State Key Laboratory of Surface Physics and Department of Physics, Fudan University, Shanghai 200433, China}

\author{Xianggang Qiu}
\affiliation{Beijing National Laboratory for Condensed Matter Physics, Institute of Physics, Chinese Academy of Sciences, P.O. Box 603, Beijing 100190, China}
\affiliation{School of Physical Sciences, University of Chinese Academy of Sciences, Beijing 100049, China}

\author{Zhiping Yin}
\email[]{yinzhiping@bnu.edu.cn}
\affiliation{Center for Advanced Quantum Studies and Department of Physics, Beijing Normal University, Beijing 100875, China}

\author{Christian Bernhard}
\email[]{christian.bernhard@unifr.ch}
\affiliation{University of Fribourg, Department of Physics and Fribourg Center for Nanomaterials, Chemin du Mus\'{e}e 3, CH-1700 Fribourg, Switzerland}

\date{\today}

%
%
\begin{abstract}
The heavy fermion (HF) state of $d$-electron systems is of great current interest since it exhibits various exotic phases and phenomena that are reminiscent of the Kondo effect in $f$-electron HF systems. Here, we present a combined infrared spectroscopy and first-principles band structure calculation study of the $3d$-electron HF compound YFe$_2$Ge$_2$. The infrared response exhibits several charge-dynamical hallmarks of HF and a corresponding scaling behavior that resemble those of the $f$-electron HF systems. In particular, the low-temperature spectra reveal a dramatic narrowing of the Drude response along with the appearance of a hybridization gap ($\Delta \sim$ 50 meV) and a strongly enhanced quasiparticle effective mass. Moreover, the temperature dependence of the infrared response indicates a crossover around $T^{\ast} \sim$ 100 K from a coherent state at low temperature to a quasi-incoherent one at high temperature. Despite of these striking similarities, our band structure calculations suggest that the mechanism underlying the HF behavior in YFe$_2$Ge$_2$ is distinct from the Kondo scenario of the $f$-electron HF compounds and even from that of the $d$-electron iron-arsenide superconductor KFe$_2$As$_2$. For the latter, the HF state is driven by orbital-selective correlations due to a strong Hund's coupling. Instead, for YFe$_2$Ge$_2$ the HF behavior originates from the band flatness near the Fermi level induced by the combined effects of kinetic frustration from a destructive interference between the direct Fe-Fe and indirect Fe-Ge-Fe hoppings, band hybridization involving Fe $3d$ and Y $4d$ electrons, and electron correlations. This highlights that rather different mechanisms can be at the heart of the HF state in $d$-electron systems.
\end{abstract}

\maketitle

%
%
The electronic band structure of solids plays a crucial role in determining their physical properties. Flat bands, characterized by a lack of dispersion over sizeable momentum ranges, are particularly intriguing. They typically give rise to an extremely singular density of states (DOS) and super-heavy electrons that tend to get localized as the electron-electron Coulomb interaction dominates over the quenched kinetic energy. This provides a fundamental platform for realizing a variety of quantum phenomena~\cite{Lieb1989,Sharpe2019,Stewart2001RMP,Hofmann202PRL,Regnault2011PRX,Neupert2011PRL,Xie2021Nat,Cao2018Natb,Cao2018Nat}, including magnetism, Mott insulators, density waves, non-Fermi liquid behavior, fractional quantum Hall effect, and unconventional superconductivity. For instance, in moir\'{e} materials, like twisted bilayer graphene (TBG), the flat bands can be shaped via the twist angle to obtain correlated insulator states and strong-coupling superconductivity~\cite{Cao2018Natb,Cao2018Nat} for which the phase diagram resembles that of the high-$T_c$ cuprates. Likewise, In a geometrically frustrated kagome lattice, the coexistence of Dirac crossings and flat bands offers an intriguing opportunity to explore novel physics involving both correlation and topology~\cite{Yin2022Nat}. Prominent examples are the recently studied kagome compounds Co$_3$Sn$_2$S$_2$~\cite{Liu2018NP,Yin2019NP}, AV$_3$Sb$_5$ (A $=$ K, Rb, Cs)~\cite{Ortiz2020PRL,Chen2021Nat,Nie2022Nat}, TbMn$_6$Sn$_6$~\cite{Yin2020Nat}, CoSn~\cite{Kang2020NC,Liu2020NC} and FeGe~\cite{Teng2022NC}. Another prominent example are heavy fermion (HF) materials~\cite{Stewart1984}, where a flat band in the vicinity of the Fermi level is achieved via the Kondo hybridization between itinerant conduction electrons and localized $f$ electrons.

Beyond the $f$-electron HF materials, there is a growing interest in exploring flat bands and HF states in $d$-electron systems, including compounds like CaCu$_3$Ir$_4$O$_{12}$~\cite{Cheng2013PRL}, CaCu$_3$Ru$_4$O$_{12}$~\cite{Kobayashi2004,Liu2020PRB}, Ca$_{2-x}$Sr$_x$RuO$_4$~\cite{Kim2022}, LiV$_2$O$_4$~\cite{Kondo1997PRL}, Fe$_3$GeTe$_2$~\cite{Zhang2018}, and iron-based superconductors (FeSCs)~\cite{Wu2016PRL,Kim2023,DeMedici2011,Georges2013,DeMedici2014}. This interest stems from the multi-orbital nature of $d$-electrons and their orbital-selective renormalization. In these systems, the electronic correlations, especially Hund's coupling, have different influences on the bands, leading to an enhanced orbital differentiation~\cite{DeMedici2011,Georges2013,DeMedici2014}. Consequently, the $d$-electron HF-type state can be achieved by doping a Hund's metal with pronounced orbital-selective correlations towards half-filling, where carriers on certain orbitals become localized while others remain itinerant, thus resembling the coexistence of light and heavy electrons originating from $s$ and $f$ orbitals, respectively, in $f$-electron HF systems. Of particular interest are the heavily hole doped FeSCs AFe$_2$As$_2$(A $=$ K, Rb, Cs) of the so-called 122 family. For instance, KFe$_2$As$_2$ exhibits a remarkable mass enhancement with a large Sommerfeld coefficient, $\gamma \sim$ 100 mJ$\cdot$mol$^{-1}$K$^{-2}$~\cite{Kim2011PRB}, comparable to that of $f$-electron HF materials. Quantum oscillation and angle-resolved photoemission spectroscopy (ARPES) experiments also support a strong mass enhancement~\cite{Terashima2010,Terashima2013,Sato2009PRL}. Additionally, a coherence-incoherence crossover, similar to that of $f$-electron HFs, has been observed in KFe$_2$As$_2$~\cite{Hardy2013,Wu2016PRL,Wiecki2018}. The coexistence of itinerant and local characters of the Fe 3$d$ electrons, along with the possible interplay of orbital-selective Hund and Kondo physics, make FeSCs a unique paradigm for exploring rich emergent quantum many-body phenomena in $d$-electron HF systems.

YFe$_2$Ge$_2$ is the parent compound of a new class of FeSCs, the so-called iron germanides, that is formally isoelectronic to KFe$_2$As$_2$ but has a so-called collapsed crystal structure, similar to that of KFe$_2$As$_2$ under high pressure~\cite{Singh2014PRB,Chen2016}. YFe$_2$Ge$_2$ is thus an interesting reference compound to KFe$_2$As$_2$ which may allow one to study the essential features of the $d$-electron HF phenomenon. In YFe$_2$Ge$_2$, the resistivity indicates a breakdown of Fermi liquid behavior at low temperatures~\cite{Zou2014,Chen2016}, and unconventional superconductivity~\cite{Subedi2014,Chen2016,Chen2019,Chen2020}, possibly on the verge of spin-triplet pairing~\cite{Zhao2020PRB}, is confirmed at $T_c \sim 1.8$~K. YFe$_2$Ge$_2$ is paramagnetic at room temperature with no magnetic phase transition down to the lowest measured temperature. However, large fluctuating magnetic moments have been observed with X-ray photoemission spectroscopy~\cite{Sirica2015} and spin susceptibility measurements~\cite{Ferstl2006}. Inelastic neutron scattering experiments further revealed the coexistence of anisotropic stripe-type antiferromagnetic (AFM) and isotropic ferromagnetic (FM) spin fluctuations~\cite{Wo2019}. Recent nuclear magnetic resonance (NMR) experiments have suggested that YFe$_2$Ge$_2$ is likely close to an itinerant magnetic quantum critical point~\cite{Zhao2020PRB}. With respect to the HF properties, an unusually high Sommerfeld coefficient $\gamma \sim$ 100 mJ$\cdot$mol$^{-1}$K$^{-2}$ has been measured with specific heat in YFe$_2$Ge$_2$~\cite{Chen2016,Chen2020}. A strong mass enhancement has been also supported by recent quantum oscillation and ARPES experiments~\cite{Baglo2022,Kurleto2023}. A coherence-incoherence crossover has been revealed and attributed to Hund's coupling induced electronic correlations~\cite{Zhao2020PRB}. In addition, a flat band feature near the Fermi level has been observed in the ARPES measurements~\cite{Xu2016PRB,Kurleto2023}. Regarding the origin of these HF phenomena, there exists no consensus amongst the various experimental reports. In some cases, they have been attributed to possible Kondo physics~\cite{Kurleto2023}, while in others they have been interpreted in terms of the orbital-differentiation physics in the Hund's limit~\cite{Zhao2020PRB,Bosse2016}. This urgently calls for further combined experimental and theoretical studies to clarify the origin of the unusual HF-like properties of YFe$_2$Ge$_2$.

\begin{figure*}[t]
\includegraphics[width=2\columnwidth]{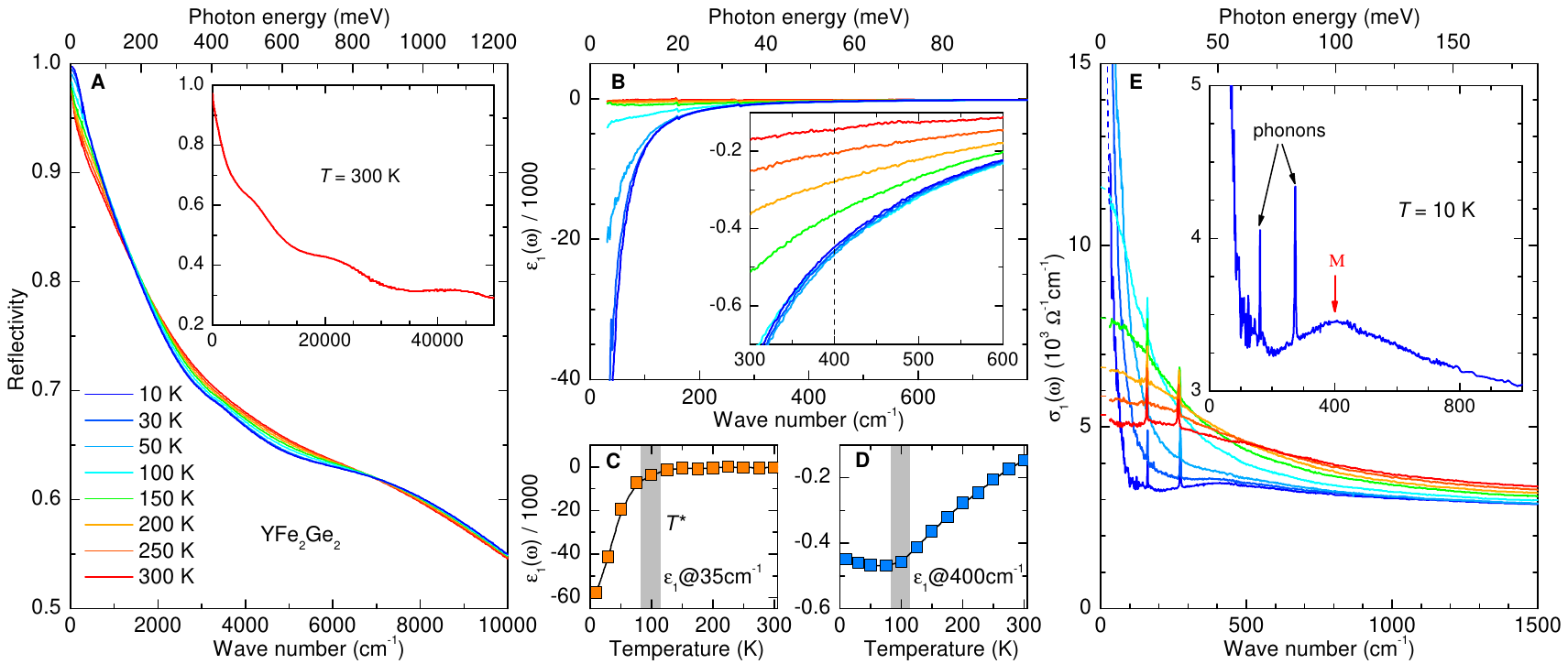}
\caption{(color online) (A) Temperature-dependent reflectivity spectra of YFe$_2$Ge$_2$ in the infrared range up to 10\,000\icm. Inset: Reflectivity spectrum at 300 K up to 50\,000\icm. (B) Temperature dependence of the real part of the dielectric function, $\varepsilon_1(\omega)$. Inset: Enlarged view of $\varepsilon_1(\omega)$ from 300 to 600\icm. (C) Temperature dependence of $\varepsilon_1(\omega)$ at 35\icm\ showing a pronounced slope change due to a coherence-incoherence crossover around $T^{\ast} \sim$ 100 K (gray bar). (D) Corresponding temperature dependence of $\varepsilon_1(\omega)$ at 400\icm. (E) Temperature-dependent spectra of the optical conductivity, $\sigma_1(\omega)$, up to 1\,500\icm\ revealing a narrow Drude peak and a low-energy band around 400\icm\ that emerges below $T^{\ast} \sim$ 100 K. Dashed lines show extrapolations to zero frequency. Inset: Magnified view of the 10 K spectrum highlighting the very narrow Drude peak, the band around 400\icm\ (M) and two infrared-active phonons.}
\label{Fig1}
\end{figure*}
%

\begin{figure}[t]
\includegraphics[width=\columnwidth]{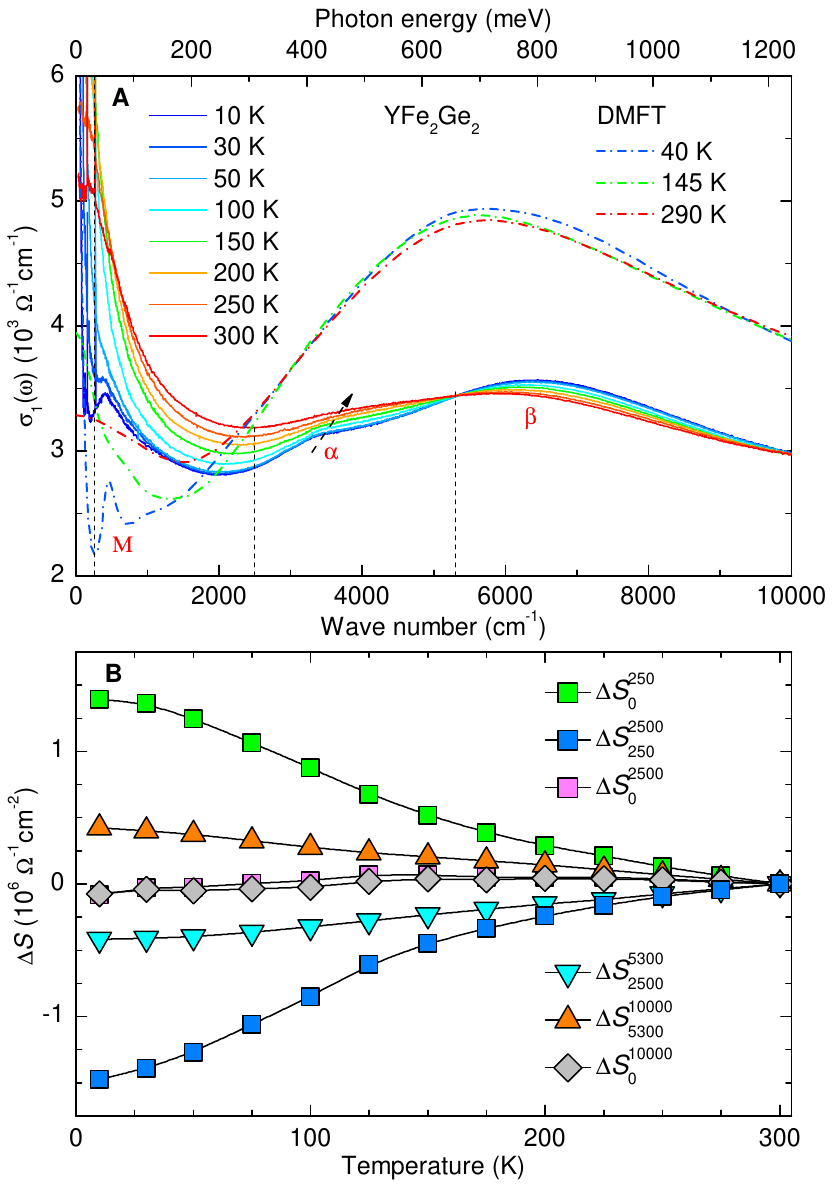}
\caption{(color online) (A) Temperature-dependent spectra of the optical conductivity, $\sigma_1(\omega)$, in the infrared range up to 10\,000\icm. Dashed lines show the conductivity obtained from DFT+DMFT calculations. (B) Temperature-dependence of the spectral weight changes, $\Delta S(T) = S(T) - S(300~\mathrm{K})$, for different frequency ranges with the cutoff frequencies marked by the vertical dashed lines in panel(A).}
\label{Fig2}
\end{figure}
In this work, we address this problem with combined infrared spectroscopy and first-principles band structure calculation studies of YFe$_2$Ge$_2$. In the infrared response, we identify all the typical features of a $d$-electron HF system, including a very narrow Drude peak due to charge carriers with a strongly enhanced effective mass, a hybridization gap, an electronic crossover from a coherent low temperature to a quasi-incoherent high temperature state, and a characteristic scaling behavior. Nevertheless, our theoretical calculations reveal a distinct mechanism for the emergence of a flat band at the Fermi level and a subsequent $d$-electron HF state in YFe$_2$Ge$_2$. Here, the source of band flatness is primarily from a kinetic frustration due to a destructive interference effect between the direct Fe-Fe and indirect Fe-Ge-Fe hopping channels which for the collapsed structure of YFe$_2$Ge$_2$ have a similar magnitude but opposite signs. An important role is also played by the band hybridization involving Fe $3d$ and Y $4d$ electrons, whereas the Hund's coupling and the related orbital differentation of the electron correlations appear to be secondary effects.

%
%
High-quality single crystals of YFe$_2$Ge$_2$ were synthesized using the Sn-flux method~\cite{Wo2019}. Details of the sample synthesis and experimental methods are included in the supplemental materials.

%

Figure 1A shows the temperature dependence of the in-plane reflectivity of YFe$_2$Ge$_2$ in the infrared region below 10\,000\icm. The inset displays the room-temperature spectrum over the full measured range up to 50\,000\icm. The overall shape of the reflectivity curves, with a zero-frequency value close to unity, signals a metallic response with a moderate screened plasma frequency of about 3\,000\icm. In the following we discuss the electronic response and its changes with temperature in terms of the dielectric function and the related optical conductivity that have been derived from the reflectivity spectra.

Figure 1B displays the temperature-dependent spectra of the real part of the dielectric function, $\varepsilon_1(\omega)$, in the far-infrared region. The overall negative values of $\varepsilon_1(\omega)$ are characteristic of the inductive response of a metal. Notably, at frequencies below about 200\icm, the spectra undergo some drastic changes at low temperatures. Whereas above 100 K, $\varepsilon_1(\omega)$ levels off at low frequencies, implying a rather large scattering rate, below 100 K it exhibits a sharp decrease with $\varepsilon_1(\omega) \sim -1/\omega^2$, which signifies a major reduction of the scattering rate. This characteristic behavior indicates a temperature-induced crossover of the charge dynamics from a high-temperature quasi incoherent to a low-temperature coherent state. As shown in Fig. 1C, the crossover temperature $T^{\ast} \sim$ 100 K can be readily determined from the sudden slope change in the temperature dependence of the low-frequency value $\varepsilon_1(35\icm)$, as marked by a gray bar. Figure 1D displays the corresponding temperature dependence of $\varepsilon_1(400\icm)$ which shows that the sudden decrease of $\varepsilon_1(\omega)$ below $T^{\ast} \sim$ 100 K occurs only at very low frequencies, meaning that the screened plasma frequency of the weakly scattered carriers is less than 400\icm. The value of $\varepsilon_1(400\icm)$ exhibits instead a weak upturn below $T^{\ast}$ which arises from a new low-energy electronic band that emerges in the low-temperature coherent state. This band with a maximum around 400\icm\ is also seen in the corresponding spectra of the real part of the optical conductivity, $\sigma_1(\omega)$, in Fig. 1E, where it is marked in the inset by a red arrow (M). The latter also details two sharp infrared-active phonon modes around 160\icm\ and 270\icm.

The $\sigma_1(\omega)$ spectra in Fig. 1E confirm the above described strong narrowing of the Drude response below 100 K. At high temperature they reveal a rather broad Drude peak with a strong tail toward high frequencies, due to intraband excitations of carriers with a large scattering rate that are quasi incoherent. Toward lower temperature, the Drude peak shows a pronounced narrowing effect that signals a strong reduction of the scattering rate, i.e., its low-frequency head gets strongly enhanced while the high-frequency tail decreases correspondingly. In particular, below $T^{\ast} \sim$ 100 K the Drude peak becomes very sharp and resembles that typically observed in HF systems~\cite{Chen2016RPP}. The infrared spectra thus signal a rather sudden change in the charge dynamics of YFe$_2$Ge$_2$ from a quasi incoherent high-temperature state with a large scattering rate, to a coherent state with a very small scattering rate below $T^{\ast}$.

Figure 2A displays the temperature-dependent $\sigma_1(\omega)$ spectra for a wider frequency range up to 10\,000\icm, which includes two pronounced interband transitions. The latter give rise to bands with maxima around 3\,500\icm\ and 6\,500\icm, which in the following are denoted as $\alpha$ and $\beta$ bands. This double-peak structure also shows a substantial variation with temperature. With increasing temperature, the $\alpha$ and $\beta$ bands both become broader. However, whereas the $\alpha$ band moves to slightly higher energy, the $\beta$ band is shifted towards lower energy. Likewise, the spectral weight of the $\alpha$ peak increases, while that of the $\beta$ peak decreases. These trends lead to a gradual blurring of the double-peak structure at elevated temperatures.

\begin{figure}[tb]
\includegraphics[width=\linewidth]{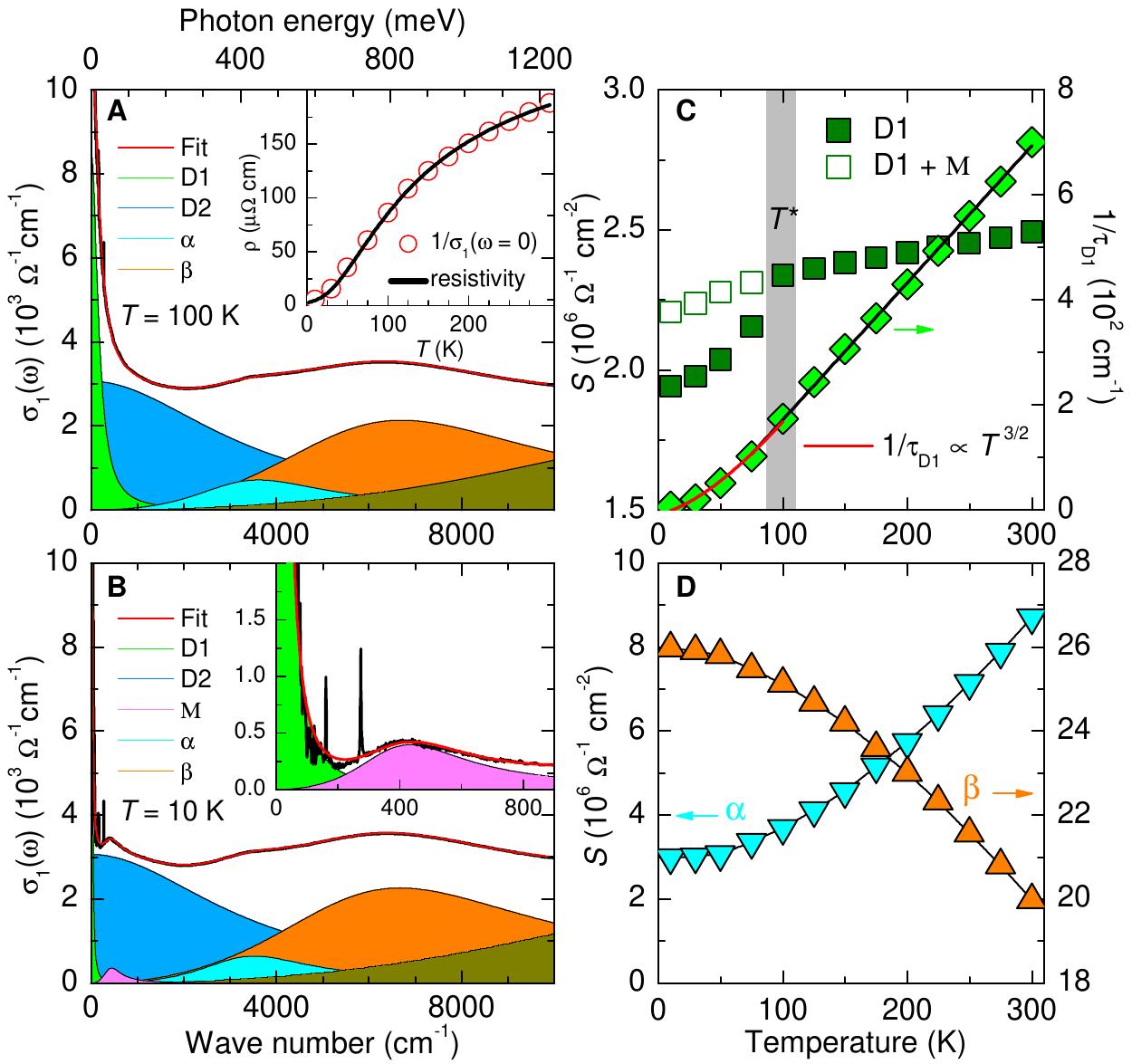}
\caption{ (color online) Decomposition of the optical conductivity spectra of YFe$_2$Ge$_2$ at (A) $T = 100$ K and (B) $T = 10$ K using a Drude-Lorentz model. The inset of panel (A) shows a comparison of 1/$\sigma_1(\omega \rightarrow 0)$, calculated from the Drude-parameters (red circles), and the dc resistivity $\rho$ from transport measurement (black line). The inset of panel (B) shows the low-frequency region after subtracting the contribution of the broad Drude peak (D2) to highlight the features of the narrow Drude peak (D1) and the low-energy peak (M). (C) Temperature dependence of the fit parameters of the narrow Drude peak (D1). The solid squares show the spectral weight of D1, and the open squares show the combined spectral weight of the D1 and M peaks. The solid diamonds represent the scattering rate of D1. The solid red line shows a fit with a $T^{3/2}$ scaling at low temperatures. (D) Temperature-dependent changes of the spectral weight of the $\alpha$ and $\beta$ bands.}
\label{Fig3}
\end{figure}

The above described temperature-dependent changes of the electronic response have been further analyzed in terms of the partial spectral weight, $S_{\omega_a}^{\omega_b}(T) = \int_{\omega_a}^{\omega_b}\sigma_1(\omega, T)d\omega$, within certain frequency ranges as defined by the lower and upper cutoff frequencies $\omega_a$ and $\omega_b$, respectively. For suitable choices of $\omega_a$ and $\omega_b$, as shown by the dashed lines in Fig. 2A, this allows us to specify the spectral weight changes of the different electronic excitations, i.e., of the Drude peak in the ranges from 0 to 250\icm\ and from 250 to 2\,500\icm, of the $\alpha$ band from 2\,500 to 5\,300\icm, and of the $\beta$ band between 5\,300 and 10\,000\icm. Figure 2B details the spectral weight changes with respect to the room temperature value, $\Delta S(T) = S(T) - S(300~\mathrm{K})$, for the various cutoff frequencies. In the ranges of 0 -- 250\icm\ and 250 -- 2\,500\icm, that are governed by the coherent and incoherent excitations of the free carriers, respectively, the incoherence-coherence crossover leads to an increase of $\Delta S_{0}^{250}$ and a corresponding decrease of $\Delta S_{250}^{2500}$. Notably, the reduction in $\Delta S_{250}^{2500}$ almost matches the enhancement in $\Delta S_{0}^{250}$, as confirmed by the almost temperature-independent value of $\Delta S_{0}^{2500}$. This conservation of the partial spectral weight below 2\,500\icm\ confirms that this low-energy range is governed by a Drude-type response with a rather broad and strongly temperature-dependent tail. Correspondingly, for the frequency ranges of 2\,500 -- 5\,300\icm\ and 5\,300 -- 10\,000\icm, that are dominated by the $\alpha$ and $\beta$ excitations, respectively, it is evident that the decrease of $\Delta S_{2500}^{5300}$ is almost compensated by the rise of $\Delta S_{5300}^{10000}$, such that $\Delta S_{2500}^{10000}$ and $\Delta S_{0}^{10000}$ remain almost constant.

\begin{figure}[tb]
\includegraphics[width=\linewidth]{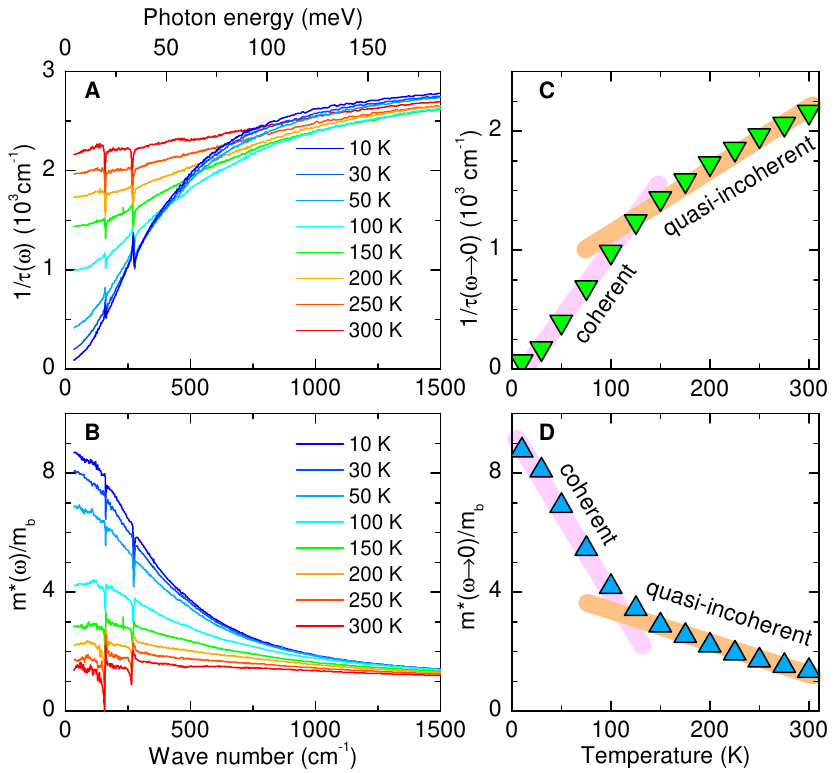}
\caption{ (color online) (A) and (B) Spectra of the frequency-dependent scattering rate $1/\tau(\omega)$ and mass enhancement $m^{\ast}(\omega)/m_b$, respectively, obtained with an extended Drude model. (C) and (D) Temperature dependence of the zero-frequency values $1/\tau(\omega \rightarrow 0)$ and $m^{\ast}(\omega \rightarrow 0)/m_b$, respectively. The colored bars indicate the slope changes around $T^{\ast} \sim$ 100 K.}
\label{Fig4}
\end{figure}

A quantitative analysis of the temperature evolution of the various intra- and interband excitations, has been obtained with a Drude-Lorentz model fit of the $\sigma_1(\omega)$ spectra. A comprehensive description of the Drude-Lorentz model is provided in Supporting information, section B. Figures 3A and 3B detail the decomposition of the $\sigma_1(\omega)$ spectra at $T =$ 100 K and 10 K, respectively, with the contribution of the various electronic bands shown in different colors. It confirms that the low-energy part of the spectra is well accounted for by two Drude components: a narrow one (D1 in green) with a small scattering rate and a broad one (D2 in blue) with a very large scattering rate. At $T= 100$ K these scattering rates amount to $1/\tau \sim$ 174\icm\ and 3\,400\icm, respectively. Such a two-Drude analysis has also been successfully used to describe the optical response of the iron-based superconductors and of other multiband systems~\cite{Wu2010,Dai2013PRL,Xu2017}. The broad component typically accounts for a temperature-independent background from nearly incoherent excitations. The temperature dependence of the low-frequency optical response is therefore governed by the narrow Drude component. In analogy, we have first fitted the weight and the scattering rate of the broad D2 component for the spectrum at 100 K and then fixed these parameters for fitting the spectra at lower and higher temperatures. The obtained temperature dependence of the parameters of the D1 component is presented in Fig. 3C. Above $T^{\ast}$, the value of $1/\tau_{D1}$ decreases linearly with decreasing temperature, whereas below $T^{\ast}$ it follows approximately a $T^{3/2}$ power-law that signals a breakdown of the Fermi liquid behavior. A similar low-temperature power-law behavior occurs in the resistivity data that are displayed in the inset of Fig. 3A (solid black line). Also shown, for comparison, are the values of the dc resistivity, $\rho \equiv 1/\sigma_1(\omega \rightarrow 0)$, calculated from the fit parameters of the two Drude components (open circles), which agree rather well with those from the transport measurements. The spectral weight of the D1 peak shows a gradual decrease towards low temperature that is anomalously enhanced below $T^{\ast} \sim$ 100 K. The additional spectra weight loss of the D1 peak can be attributed to the emergence of the M band below $T^{\ast}$ which apparently develops at the expense of the D1 peak. This M band has been described by a weak Lorentz band, as shown by the magenta line and shading in Fig. 3B. The open squares in Fig. 3C confirm that the combined spectral weight of the D1 and M bands does indeed not show a corresponding anomaly in the vicinity of $T^{\ast} \sim$ 100 K.

\begin{figure*}[tb]
\centering
\includegraphics[width=0.9\textwidth]{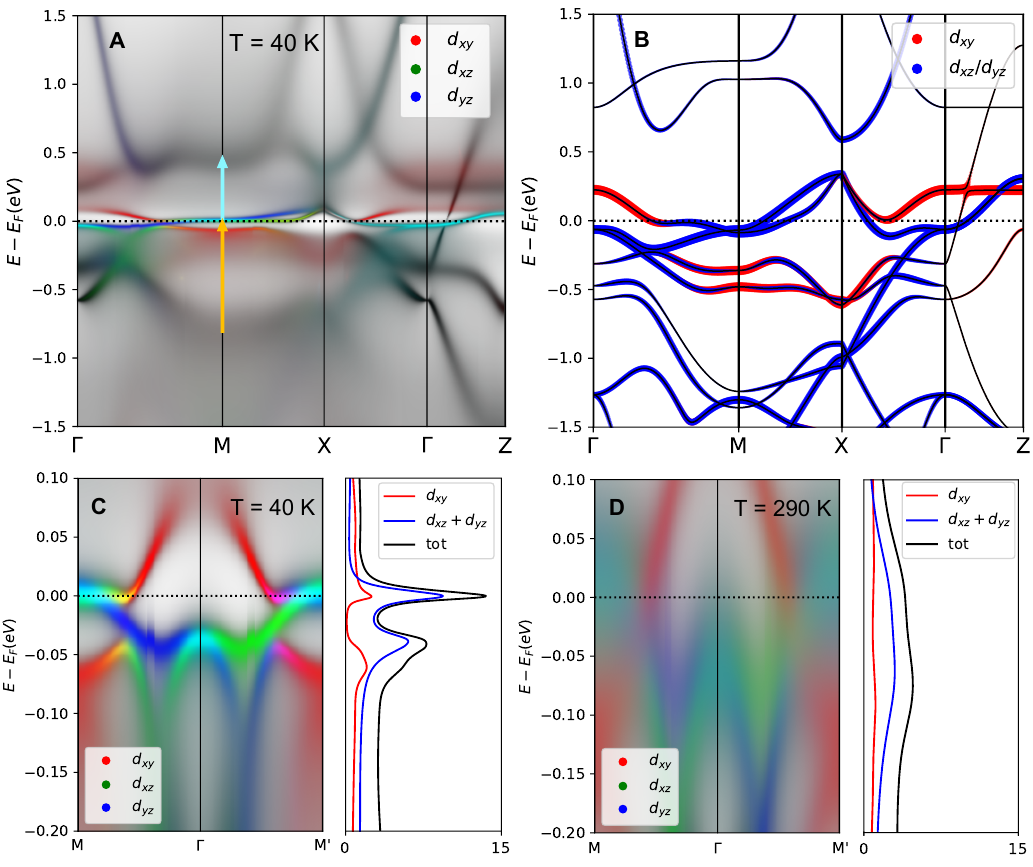}
\caption{ (color online) (A) Orbital-resolved electronic band structure of YFe$_2$Ge$_2$ obtained by the DFT+DMFT method at $T = 40$ K. Arrows indicate the corresponding optical transitions for the $\alpha$ and $\beta$ bands in the infrared spectra. (B) Orbital-resolved electronic band structure calculated with the DFT method, which neglects electronic correlations. (C) and (D) Orbital-resolved electronic structure and DOS along the $\Gamma$--M/M$^{\prime}$ directions near the Fermi level obtained by the DFT+DMFT method at $T =$ 40 K and 290 K, respectively.}
\label{Fig5}
\end{figure*}

The inset of Fig. 3B shows the far-infrared response at 10 K with the contribution of the broad Drude peak subtracted. It highlights that the response of the extremely narrow D1 band (with $1/\tau \sim$ 5\icm\ $< k_B T \sim 7\icm$) and the emerging M band are indeed highly reminiscent of the optical response of a HF system~\cite{Chen2016RPP,Singley2002}. In the latter case, these two components correspond to the intraband response of the heavy quasiparticles and the excitations across the Kondo hybridization gap ($\Delta$), respectively. In analogy, in the following we assign the M-mode to excitations across a hybridization gap.

The double-peak structure associated with the $\alpha$ and $\beta$ interband transitions at higher energies has been described with two additional Lorentz terms. Figure 3D details their spectral weight changes with temperature and shows that they have opposite trends and thus nearly cancel each other. This confirms that the high energy spectral weight redistribution occurs predominantly between these $\alpha$ and $\beta$ bands.

\begin{figure}[tb]
\centering
\includegraphics[width=0.95\linewidth]{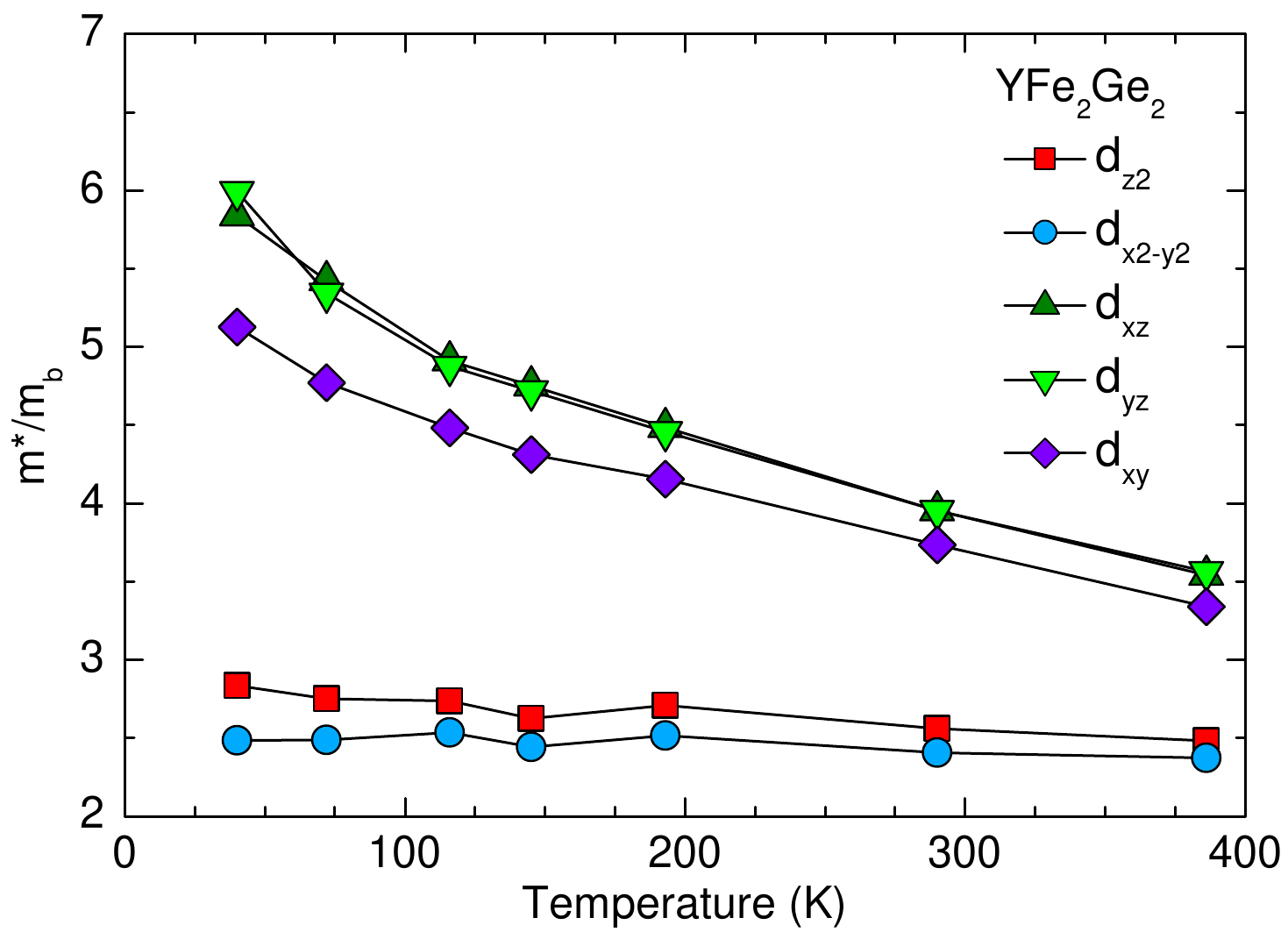}
\caption{ (color online) Temperature dependence of the orbital-resolved mass enhancement for YFe$_2$Ge$_2$ obtained from the DFT+DMFT calculations.}
\label{Fig6}
\end{figure}
An alternative approach to analyze the low-energy electronic response involves the so-called extended Drude model (EDM). It is typically employed to quantify the renormalization due to electronic correlations in strongly correlated materials. Here a frequency-dependent mass enhancement $m^\ast(\omega)/m_b$ and scattering rate $1/\tau(\omega)$ are derived from the measured infrared spectra. Further details about the EDM analysis are provided in the Supporting information, section C. The obtained spectra of $1/\tau(\omega)$ and $m^\ast(\omega)/m_b$ are displayed in Figs. 4A and 4B, respectively. The temperature dependence of the extrapolated zero-frequency values $1/\tau(\omega \rightarrow 0)$ and $m^\ast(\omega \rightarrow 0)/m_b$ is shown in Figs. 4C and 4D, respectively. The scattering rate $1/\tau(\omega \rightarrow 0)$ exhibits a strong decrease below $T^{\ast} \sim$ 100 K, which confirms the emergence of a coherent low-temperature state. The corresponding spectra of the effective mass in Fig. 4B, reveal a sizeable enhancement that becomes very prominent below $T^{\ast} \sim$ 100 K. The zero-frequency value $m^\ast(\omega \rightarrow 0)/m_b$ in Fig. 4D exhibits a clear anomaly below $T^{\ast} \sim$ 100 K, where it increases more rapidly than above $T^{\ast}$ reaching up to $m^\ast/m_b \sim 10$. This increase signals a sizeable renormalization of (some of) the electronic bands in YFe$_2$Ge$_2$ that is reminiscent of the behavior found in other correlated metals such as the HF compounds~\cite{Dordevic2001PRL,Basov2011}. Note, that this analysis of the optical response only yields an average of the effective mass of the various bands in the vicinity of the Fermi-level which, in case of a strong orbital differentation, may underestimate the effective mass of a particular band.

To better understand the charge dynamics of YFe$_2$Ge$_2$, we have also performed first-principles calculations combining density functional theory (DFT) and dynamical mean-field theory (DMFT). The DFT+DMFT technique has successfully described the electronic structures of many iron-based superconductors~\cite{Yin2011NM,Yin2011NP,Yin2014NP}. Figure 5A shows the DFT+DMFT band structure at $T = 40$ K. The first remarkable observation concerns the flat band feature at the Fermi level that extends over a fairly wide momentum range. This flat band is dominated by the $d_{xz/yz}$ orbital character around the M point in the two-iron Brillouin zone representation. As detailed in the Supporting information, this flat band region constitutes about 15.6\% of the first Brillouin zone. It is noteworthy that recent ARPES measurements have also identified such a flat band feature within a few meV of the Fermi level~\cite{Xu2016PRB,Kurleto2023}. Another important observation concerns the small hybridization gap between the $d_{xz/yz}$ band and the $d_{xy}$ band that is also in close proximity to the Fermi level. As highlighted in Fig. 5C, the size of this hybridization gap amounts to $\Delta \sim$ 50 meV. Fig. 5C and Fig. 5D compare the band structure and the density of states (DOS) of the Fe $3d$ orbitals along the $\Gamma$--M/M$^{\prime}$ directions at $T = 40$ K and 290 K, respectively. Additional calculations for different temperatures can be found in the Supporting information, section D. At $T = 40$ K, owing to the flat $d_{xz/yz}$ band, the quasi-particle peak at the Fermi level is sharply defined and resembles that of a so-called Kondo resonance peak. Upon increasing temperature, the quasi-particle peak moves slightly down in energy and undergoes a strong broadening that signals a coherence-incoherence crossover. Note, however, that even at $T = 290$ K the hybridization gap has not entirely vanished, as evidenced by the slight suppression of the DOS in the gap region, and is only smeared out by the band broadening induced by the strong thermal excitations and related effects.

\begin{figure*}[tb]
\centering
\includegraphics[width=0.95\textwidth]{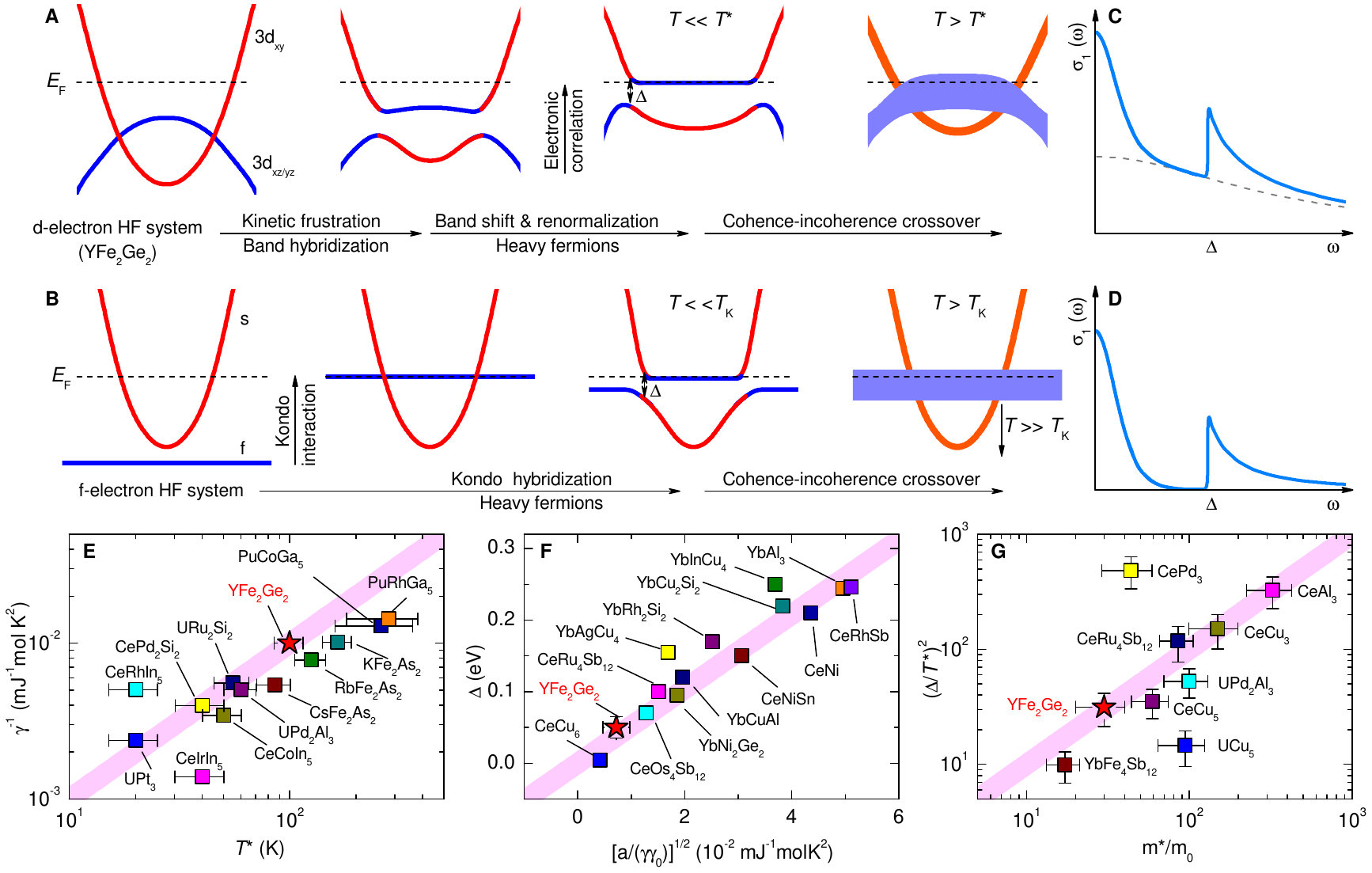}
\caption{ (color online) (A) and (B) Schematics of the development of a flat band and hybridization gap at the Fermi level in $d$-electron and $f$-electron systems, respectively. (C) and (D) Corresponding $\sigma_1(\omega)$ spectra in the HF state for the coherent Drude response of heavy quasiparticles and excitations across the hybridization gap. The dashed gray line in (C) denotes a broad Drude response from other incoherent $d$-electron bands. (E)--(G) Scaling plots of $\gamma^{-1}$ vs. $T^{\ast}$, $\Delta$ vs. [$a/(\gamma\gamma_0)$]$^{1/2}$, and $(\Delta/T^{\ast})^2$ vs. $m^{\ast}/m_0$, respectively, for various HF compounds. The red star shows the corresponding scaling behavior of YFe$_2$Ge$_2$. The values for the other HF compounds have been adopted from Refs.~\cite{Wu2016PRL,Okamura2007,Dordevic2001PRL} and references cited therein.}
\label{Fig7}
\end{figure*}
The presence of the flat band and the hybridization gap at the Fermi level naturally explain the main features of the measured infrared conductivity spectra. The calculated $\sigma_1(\omega)$ spectra, as shown in Fig. 2A, reproduce quite well the various low-energy intra- and interband excitations. Here, the narrow Drude peak (D1) arises from the intraband excitations of the flat band, while the M band originates from the excitations across the hybridization gap. The substantial temperature dependence of the spectral weight of the $\alpha$ and $\beta$ bands can also be readily understood, since the $\alpha$ band involves transitions with $\sim 0.5$ eV from the flat band to a higher energy band (cyan arrow), while the $\beta$ peak is contributed by the transitions with $\sim 0.75$ eV from a lower energy band to the flat band (orange arrow). As the temperature increases, the flat band is broadened and shifted to slightly lower energy which also explains the anomalous blue-shift of the $\alpha$ band. The consequent increase (decrease) of the joint density of states for the $\alpha$ ($\beta$) excitations naturally accounts for the observed transfer of spectral weight between these two bands. The development of a coherent state in the flat band is at the heart of the suppression of the scattering rate and the enhancement of the effective mass of the charge carriers seen in Figs. 4C and 4D. Note that the EDM analysis involves contributions from all conduction bands, and therefore yields an averaged value of the quasiparticle effective mass that can be quite a bit smaller than that of the flat band. The infrared data thus agree with recent ARPES and quantum oscillation experiments~\cite{Baglo2022,Kurleto2023}, which yield largee estimates of the quasiparticle effective mass of the flat band of $m^\ast \sim$ 25 -- 30 $m_0$ ($m_0$ is the bare electron mass).

The flat band and the hybridization gap at the Fermi level in YFe$_2$Ge$_2$ thus account for the characteristic features in the optical response that are amazingly similar to those found in Kondo-type HF systems. However, there are also indications that some aspects of the optical response of YFe$_2$Ge$_2$ cannot be explained in terms of a classical Kondo scenario. For example, the quasiparticle (Drude) peak and the hybridization gap do not vanish entirely as the temperature is increased above $T^{\ast}$, but are only blurred due to the broadening of these bands. In return, this implies that a classical Kondo hybridization scenario does not entirely explain the behavior of the flat band and the hybridization gap in YFe$_2$Ge$_2$. Moreover, it indicates that the Fe $3d_{xz/yz}$ electrons are not fully localized at high temperatures and, accordingly, this compound does not appear to be in the Kondo-limit.

Notably, even for the comparison of YFe$_2$Ge$_2$ and KFe$_2$As$_2$ it appears that, despite of their strikingly similar infrared response (details are shown in the Supporting information, section H), different mechanisms are causing the flat band formation and the related HF behavior. For KFe$_2$As$_2$, it was previously reported that the HF behavior originates from a large Hund's coupling which leads to a strong orbital differentiation of the electron correlations~\cite{Terashima2010,Terashima2013,Sato2009PRL,DeMedici2014}. Such a Hund's metal scenario is indeed feasible in case of KFe$_2$As$_2$ with an orbital occupancy of $3d^{5.5}$ that is rather close to a half-filling (at $3d^5$). In KFe$_2$As$_2$, the mass enhancement of the $d_{xy}$ orbital has been reported to reach values as high as 15 -- 20 and to exceed that of the $d_{xz/yz}$ levels by at least a factor of three~\cite{DeMedici2014}. The strong differentiation of orbital-selective correlations thus can explain the coexistence of light and heavy $d$ electrons, and thus the analogy to the classical $f$-electron HF systems, where this role is played by the $s$ and $f$ electrons.

A different scenario arises for YFe$_2$Ge$_2$, where the orbital-resolved mass enhancement and orbital-selective correlations, as obtained from our DFT+DMFT calculations and shown in Fig. 6, appear to be weaker than in KFe$_2$As$_2$. For the $d_{z^2}$ and $d_{x^2-y^2}$ orbitals, which are anyhow far away from the Fermi level, the mass enhancement is only moderate with $m^\ast/m_b \sim$ 2 -- 3 and hardly temperature dependent. The most relevant $d_{xy}$ and $d_{xz/yz}$ orbitals have rather similar mass enhancement factors that amount to about $\sim$ 3.5 at $T = 386$ K and exhibit a moderate increase to $\sim$ 5 -- 6 at low temperature. This confirms that in YFe$_2$Ge$_2$ there is only quite a weak differentiation of the mass enhancement of the relevant bands in the vicinity of the Fermi-level, which can hardly explain the flat band and HF behavior.

To gain more insight into the origin of the flat band and the related HF state in YFe$_2$Ge$_2$, we conducted a more careful comparison between YFe$_2$Ge$_2$ and KFe$_2$As$_2$. As detailed in section H of the Supporting information, a comparison of the DFT band structures of YFe$_2$Ge$_2$ and KFe$_2$As$_2$ reveals that the $d$ bands in the vicinity of the Fermi level are much narrower for the former than for the latter. The reason for the band flattening in the absence of electronic correlations in YFe$_2$Ge$_2$ is a so-called kinetic frustration effect~\cite{Yin2011NM}. This arises due to a competition between the direct Fe-Fe and the indirect Fe-Ge-Fe hopping channels for which the hopping parameters have opposite signs and thus give rise to a strongly destructive interference effect when they have comparable amplitudes, as is the case in YFe$_2$Ge$_2$ for which the collapsed 122-structure yields a stronger Fe-Ge bonding and thus an enhanced Fe-Ge-Fe hopping parameter (as compared to the uncolapsed 122-structure of KFe$_2$As$_2$). Note that, in addition to the orbital occupancy~\cite{DeMedici2014}, the anisotropy of the direct and indirect hopping parameters is also an important factor contributing to the orbital differentiation~\cite{Yin2011NM}. Therefore, the weak anisotropy in hopping parameters for YFe$_2$Ge$_2$ also accounts for the weak orbital differentiation within the $d_{xz/yz}$ and $d_{xy}$, as observed in Fig.~6.

A second crucial factor is band hybridization, as is evident from both the DFT+DMFT and the DFT band structures, shown in Figs. 5A and 5B, respectively, which reveals that the hybridization of the $d_{xz/yz}$ and $d_{xy}$ bands (and the resulting small hybridization gap) appears in both cases. As outlined in section H of the Supporting information, this hybridization is facilitated by the partial occupation and the extended spatial expansion of the Y $4d$ electrons, enabling them to overlap with Fe $3d$ electrons. Consequently, the $d$ orbitals near the Fermi level hybridize, further flattening the bands. In comparing the DFT+DMFT and the DFT results, the influence of the electron correlations, particularly of the orbital-selective ones, appears to be a secondary effect that helps to further flatten these bands and shift them closer toward the Fermi level, where they become electronically active~\cite{Mao2023PRB}.

Figures 7A shows a schematic summary of the above described scenario for obtaining a flat band and hybridization gap at the Fermi level of YFe$_2$Ge$_2$. Figure 7B displays a corresponding sketch of the classical Kondo effect. Despite their different mechanisms, the two cases yield a very similar electronic structure of the HF state, i.e., featuring a flat band and a hybridization gap at the Fermi level. Accordingly, as illustrated in Figs. 7C and 7D, the $d$-electron and $f$-electron HF states both share very similar spectral features in the optical conductivity. Notably, for the $d$-electron HF state these features are superimposed on an additional broad Drude peak due to contributions from other $d$-electron bands that remain incoherent.

Next, we address the question whether, despite of the above discussed different mechanism of the flat band and HF behavior in YFe$_2$Ge$_2$, the scaling relationships of the $f$-electron HF compounds are still obeyed. In the following we show that this is indeed the case. In the $f$-electron HF compounds the following scaling relation between the Sommerfeld coefficient $\gamma$ and the Kondo temperature $T_K$ (or the crossover temperature $T^{\ast}$) applies: $T_K = \mathrm{R log2}/\gamma \simeq 10^4/\gamma$ [mJ$^{-1}$$\cdot$mol K$^{2}$]~\cite{Onuki2013,Zhang2018,Wu2016PRL}. Figure 7E shows that this $T^{\ast} \sim 1/\gamma$ scaling is also observed in YFe$_2$Ge$_2$ with $T^{\ast} \simeq 100$ K and a value of $\gamma \simeq$ 100 mJ$\cdot$mol$^{-1}$K$^{-2}$, as reported from specific heat~\cite{Chen2020}. Additional scaling relations, that have been derived from the hybridization of the conduction and the $f$ electrons by using the mean-field approximation to the periodic Anderson model~\cite{Millis1987PRB,Chen2016RPP,Dordevic2001PRL,Okamura2007}, are $\Delta \propto \sqrt{T_{K}W}$ and $m^{\ast}/m_0 = (\Delta/T_K)^2$. Here $\Delta$ is the direct hybridization gap, $W$ the conduction electron bandwidth, and $m^{\ast}/m_0$ the effective mass of the heavy electrons. The scaling relation $\Delta \propto \sqrt{T_{K}W}$ can be rewritten as $\Delta \propto \sqrt{a/(\gamma\gamma_0)}$. This has been demonstrated by Okamura \textit{et al} for a number of Ce- and Yb-based HF compounds, where $W$ of a Ce(Yb) HF compound can be regarded as inversely proportional to $\gamma$ of the isostructural La(Lu) non-HF compound (denoted as $\gamma_0$), and $a$ is a constant which depends only on the $f$ level degeneracy $N$: $a  =$ 0.21, 0.54, and 0.59 for $N =$ 2, 6, and 8, respectively~\cite{Okamura2007,Chen2016RPP}. Figs. 7F and 7G show that YFe$_2$Ge$_2$ does indeed obey the scaling relations of $\Delta \propto \sqrt{a/(\gamma\gamma_0)}$ and $m^{\ast}/m_0 = (\Delta/T^{\ast})^2$, with the parameters $\Delta \simeq 50$ meV, $\gamma \simeq 100$ mJ$\cdot$mol$^{-1}$K$^{-2}$, $\gamma_0 \simeq 40$ mJ$\cdot$mol$^{-1}$K$^{-2}$ (from specific heat data of LaFe2Ge2~\cite{Ebihara1995,Chen2020}), $a = 0.21$ for a doublet degeneracy, $m^\ast/m_0 \simeq 30$~\cite{Baglo2022,Kurleto2023}, and $T^\ast \simeq 100$~K.

%
%

Finally, we comment on the significance of our findings. The proposed mechanism and its similarity to the Kondo scenario not only sheds new light on the origin of HF state in YFe$_2$Ge$_2$, but also offers insights to clarify its Kondo-like properties. The emergent flat band at the Fermi level may also account for various exotic properties of YFe$_2$Ge$_2$, such as the non-Fermi-liquid behavior~\cite{Kumar2021}, the coexistence of FM and AFM spin fluctuations~\cite{Wo2019,Mao2023PRB}, the large Sommerfeld coefficient at low temperatures~\cite{Xu2016PRB,Kurleto2023} (which is also reproduced by our calculations provided in section F of the Supporting information), and even the occurrence of unconventional superconductivity~\cite{Cao2023}. On the other hand, our scenario suggests a novel route to construct and manipulate flat bands through the combined interactions of kinetic frustrations, band hybridization, and electron correlations, which can improve the participation of flat bands in the low-energy physics. Furthermore, taking into account the topology of band hybridization, this scenario will be interesting for building a rich set of novel topological quantum phenomena, such as possible topological HF state or topological superconductivity~\cite{Hao2019}. In conclusion, the present work identifies YFe$_2$Ge$_2$ as a model system that can serve as a platform for investigating the $d$-electron HF physics and related exotic properties.

%
%
\begin{acknowledgments}
This work was supported by the National Natural Science Foundation of China (Grant No. 12274442) and the National Key R\&D Program of China (Grant No. 2022YFA1403901), as well as by the Swiss National Science Foundation (SNSF) through Grants No. 200020-172611 and 200021-214905. Z.P.Y. acknowledges support by the National Natural Science Foundation of China (Grant No. 12074041).
\end{acknowledgments}

%
%

%
%
%


%

\end{document}